\documentclass[twocolumn,superscriptaddress,letter]{revtex4-2}%
\usepackage{amssymb}
\usepackage{graphicx}
\usepackage{amsmath}
\usepackage{soul}
\usepackage{amsthm}
\usepackage{amsfonts}
\usepackage[T1]{fontenc}
\usepackage[latin9]{inputenc}
\usepackage{array}
\usepackage{multirow}
\usepackage{color}
\usepackage{esint}
\usepackage{bm}
\usepackage{color}
\usepackage{bbm}
\usepackage{hyperref}
\usepackage{babel}
\usepackage{titlesec}

\newcommand{\ket}[1]{|{#1}\rangle}
\newcommand{\bra}[1]{\langle{#1}|}
\newcommand{\braket}[1]{\langle{#1}\rangle}

\newcommand{\kc}{{\mathrm{kc}}}
\newcommand{\BdG}{{\mathrm{BdG}}}

\newcommand{\ssh}{{\mathrm{ssh}}}

\newcommand{\tpsi}{{\tilde{\psi}}}
\newcommand{\tE}{{\tilde{E}}}
\newcommand{\tH}{{\tilde{H}}}

\newcommand{\Tr}{{\mathrm{Tr}}}
\begin{document}
	\global\long\def\id{\mathbbm{1}}
	\global\long\def\ui{\mathbbm{i}}
	\global\long\def\ud{\mathrm{d}}
\title{Dissipation-assisted preparation of topological boundary states}
\author{Yi Peng}
\affiliation{International Quantum Academy, Shenzhen 518048, China}
\author{Chao Yang}
\affiliation{Shenzhen Institute for Quantum Science and Engineering,
	Southern University of Science and Technology, Shenzhen 518055, China}
\affiliation{International Quantum Academy, Shenzhen 518048, China}
\affiliation{Guangdong Provincial Key Laboratory of Quantum Science and Engineering,
	Southern University of Science and Technology, Shenzhen 518055, China}
\author{Haiping Hu}
\affiliation{Beijing National Laboratory for Condensed Matter Physics, Institute of Physics, Chinese Academy of Sciences, Beijing 100190, China}
\affiliation{School of Physical Sciences, University of Chinese Academy of Sciences, Beijing 100049, China}
\author{Yucheng Wang}
\email{wangyc3@sustech.edu.cn}
\affiliation{Shenzhen Institute for Quantum Science and Engineering,
	Southern University of Science and Technology, Shenzhen 518055, China}
\affiliation{International Quantum Academy, Shenzhen 518048, China}
\affiliation{Guangdong Provincial Key Laboratory of Quantum Science and Engineering,
	Southern University of Science and Technology, Shenzhen 518055, China}
%\date{\today}

\begin{abstract}
Robust states emerging at the boundaries of a system are an important hallmark of topological matter. Here, using the Su-Schrieffer-Heeger model and the Kitaev chain as examples, we study the impact of a type of experimentally realizable bond dissipation on topological systems by calculating the steady-state density matrix, and demonstrate that such dissipation applied near the system boundary can assist in preparing topological edge states of the parent Hamiltonian, irrespective of the initial state or filling. This effect stems from the matching between the phase distribution encoded in the topological edge states and the target state prepared through bond dissipation. This work provides new insights into the preparation of topological edge states, particularly in the context of Majorana zero modes.
%especially the preparation of Majorana zero modes. 
\end{abstract}
\maketitle

%\noindent{\large{\textbf{Introduction}}}\\
Topology is a key paradigm for understanding the phases of matter and has garnered significant interest from multiple fields over the past few decades~\cite{TPBS1,TPBS2,TPBS3,TPBS4,TPBS5,TPBS6,TPBS7}. 
Topologically protected edge states, one of the most important features of topological physics, are typically found at the boundary of topological insulators, superconductors, or other topologically nontrivial systems. These states are robust against local disorder and perturbations due to their topological nature, which arises from the global properties of the system~\cite{TPBS5,TPBS6,TPBS7,TPBS8}. These robust edge states have opened up new possibilities for the development of advanced technologies, such as low-power electronics and topological quantum computing. Understanding and manipulating these edge states is crucial for both advancing fundamental physics and harnessing their potential applications in future technologies. 

Previously, most studies of edge states focused on closed systems, primarily relevant to gapped systems, where the Fermi level lies exactly at the gap and the temperature is much lower than the band gap~\cite{TPBS5,TPBS6,TPBS7,TPBS8}. However, systems inevitably interact with their environment, making it both important and necessary to broaden our understanding of topological systems to encompass non-Hermitian and open quantum systems, which have garnered widespread attention in recent years~\cite{NonH1,NonH2,NonH3,NonH4,openT1,openT2,openT3,openS1,openS2,openS3,openS4,openS5,openS6,openS7,openS8,others3,others2,others1,other0,other1,other2,other3,other4,openD1,openD2,openD3,openD4,openD5,openD6,openD7,openD8,openD9,openD10,openD11}. In open systems, where the quantum state is typically a mixed state rather than a pure state, the characterization of topology differs from that in closed systems. An open quantum system can be described by a density matrix, and its evolution is determined by the Lindblad equation~\cite{Lov1,Lov2,Lov3}. Thus, there are studies on the topological properties of open systems, focusing on the topology of the Lindbladian~\cite{openT1,openT2,openT3}, non-equilibrium steady states~\cite{openS1,openS2,openS3,openS4,openS5,openS6,openS7,openS8}, or the topological features of the density matrix~\cite{openD1,openD2,openD3,openD4,openD5,openD6,openD7,openD8,openD9,openD10,openD11}. 

In this work, we study a type of experimentally realizable bond dissipation applied near the boundary of a topological system, using the Su-Schrieffer-Heeger (SSH) model~\cite{SSH} and the Kitaev chain~\cite{Kitaev} as examples. We investigate the characteristics of the density matrix of the steady state in the eigenbasis of the parent Hamiltonian without dissipation and find that the steady state has the largest weight on the boundary states for the SSH model and on the ground state for the Kitaev chain, and this weight remains unchanged when dissipation is removed. Specifically, for the Kitaev chain, we observe that the steady state occupies the ground state with a ratio close to $1$. As is well known, a smoking-gun signature of Majorana zero modes (MZMs) has yet to be observed, and our results suggest that controllable dissipation could be considered to assist in the preparation of MZMs. Unlike previous studies on the topology of open systems, we do not focus on how to characterize their topological properties. Instead, we focus on preparing the topological boundary states of the original system by using dissipation as a control mechanism, and once the system reaches the steady state, we remove the dissipation. This approach does not require cooling to the ground state, nor does it require the Fermi surface to be precisely tuned to the gap, meaning that the properties of the system's boundary states can be manifested for arbitrary filling fractions through dissipation.\\

\noindent{\large{\textbf{Models and Results}}}\\
\textbf{bond dissipation}

\noindent The time evolution of dissipative systems usually leads to mixed states, which need to be described using the density matrix $\rho$. The evolution of $\rho$ can generally be described by quantum master equations, with the most commonly used and concise being the Lindblad master equation~\cite{Lov1,Lov2,Lov3}

\begin{equation}
	\frac{\ud\rho}{\ud{}t} 
	=\mathcal{L}[\rho]
	= -i[H,\rho] 
	+ \sum_{j}(D_j\rho{}D_j^\dagger-\frac{1}{2}\{D_j^\dagger{}D_j,\rho\}),
	\label{eq_mastereq}
\end{equation}
$\mathcal{L}$ is the Lindbladian, and we will focus on the steady-state density matrix $\rho_{s}$, which corresponds to its eigenstate with a zero eigenvalue, i.e. $\mathcal{L}[\rho_s]=0$. The jump operator $D_j$ considered in Eq. \eqref{eq_mastereq} 
acts on a pair of sites $j$ and $j+\ell$, as given by~\cite{openS1,Peter1,Peter2,Peter3,Peter4,Wang1,Wang2,Marcos,Yusipov1,Yusipov2,Haga}
\begin{equation}
	D_j = \sqrt{\Gamma_j}(c_{j+\ell}^\dagger+ac_{j}^\dagger)(c_{j+\ell}-ac_{j}).
	\label{eq_bond_diss_jump}
\end{equation}
where $a=\pm 1$, $\ell=1$ or $2$, $j=1,2,\dots,L-\ell$, and $\Gamma_j$ is the dissipation strength.
This operator does not change the system's particle number, but it alters the relative phase between the pair of sites. It drives the system from an out-of-phase (in-phase) state to an in-phase (out-of-phase) state over a distance $\ell$, corresponding to $a=1$ ($a=-1$). This type of dissipation is also known as bond dissipation. It was suggested to be realizable in cold atom systems~\cite{openS1,Peter1,Peter2,Peter3,Peter4,Wang1} or in arrays of superconducting microwave resonators~\cite{Marcos}.
We will investigate how this dissipation can drive systems of interest to topological states using the SSH model and the Kitaev chain as examples, and understand the mechanism behind it. \\

\noindent\textbf{SSH model}\\
The SSH model describes a one-dimensional (1D) chain with two sites $A$ and $B$ per unit cell, where the particle hops with staggered hopping amplitudes. The Hamiltonian is given by~\cite{SSH}
\begin{equation}\label{ssh_h0}
	{H}_\ssh 
	=\sum_{j}\left(v{c}_{jA}^\dagger{c}_{jB}+w{c}_{j+1,A}^\dag{c}_{jB}\right)+{h.c.},
\end{equation}
where $v$ and $w$ are the intra-cell and inter-cell hopping amplitudes, respectively. 
The ratio $v/w$ controls the topological phase transition: $v>w$ corresponds to the topologically trivial phase, while 
$v<w$ corresponds to the topologically non-trivial phase. In the following calculation, we set $v=1$ as the energy unit, the number of cells as $N$, and use open boundary conditions (OBC).

\begin{figure}[ht!]
	\centering
	\includegraphics[width=0.5\textwidth]{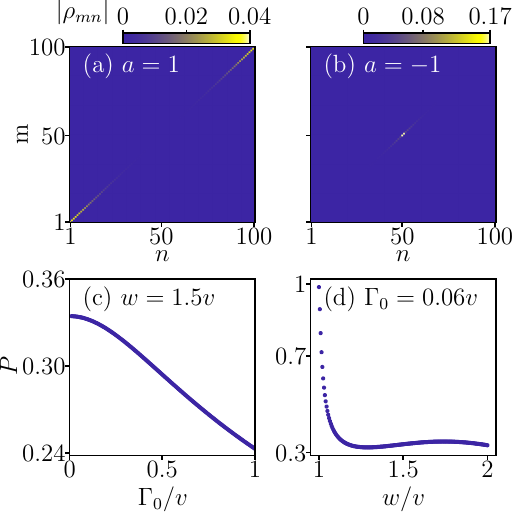}
	\caption{ In the eigenbasis of the parent Hamiltonian $H_\ssh$, the absolute values of the elements of the density matrix for steady states  with the bond dissipation (a) $a=1$ and (b) $a=-1$.
	Fixing $a=-1$, the occupation probability $P$ of edge states in the steady state as a function of 
	(c) dissipation strength $\Gamma_0$ and (d) inter-cell 
		hopping amplitude $w$. Here we fix $L=100$.}
	\label{fig1}
\end{figure}

The bond dissipation that we consider adding to the SSH model connects sites of the same sublattice, which is equivalent to 
$\ell=2$ in Eq. \eqref{eq_bond_diss_jump}, i.e.,
\begin{equation}
	D_{jX} 
	= \sqrt{\Gamma_{jX}}[c_{(j+1)X}^\dagger+ac_{jX}^\dagger][c_{(j+1)X}-ac_{jX}],
	\label{eq_ssh_diss_samesub}
\end{equation}
where $X=A,B$ and $j=1,2,\ldots,N-1$.  Furthermore, since the distribution of the edge states decays exponentially from the ends of the chain towards the bulk, we set the dissipation strength to also follow an exponential decay starting from the boundaries, i.e.,
\begin{equation}
	\Gamma_{jA} = \Gamma_0e^{-(j-1)}, \quad
	\Gamma_{jB} = \Gamma_0e^{-(N-1-j)}.
	\label{eq_ssh_diss_strength}
\end{equation}
%In the main text, we set the system size $L=100$, $w=1.5v$ for $H_\ssh$, and $\ell=1$ and $\Gamma_0=0.06v$ for bond dissipation, unless otherwise stated.

We first analyze the steady-state density matrix in the eigenbasis of the parent Hamiltonian $H_{\ssh}$, i.e., $\rho_{mn}=\langle \psi^{(m)}|\rho_{s}|\psi^{(n)}\rangle$, with $|\psi^{(m)}\rangle$ and $|\psi^{(n)}\rangle$ being the eigenstates of $H_{\ssh}$, as shown in Fig.~\ref{fig1}(a,b). Here we set the system size to $L=2N=100$, with $\Gamma_0=0.06v$ and $w=1.5v$, indicating that the system is in the topological phase. We observe that when $a=1$, the bond dissipation drives the system towards the bulk states at the edges of the energy spectrum of $H_{\ssh}$ [Fig.~\ref{fig1}(a)]. In contrast, when 
$a=-1$, the steady-state is predominantly composed of the two topological edge states located at the center of the energy spectrum of $H_{\ssh}$ [Fig.~\ref{fig1}(b)]. Numerical simulations show that the total occupation probability 
$P$ of the edge states reaches approximately $0.334$. Furthermore, we numerically find that 
$P$ changes with system size according to the relation $P=P_{\ssh}+\alpha_{\ssh}/L$, where $P_{\ssh}=0.131\pm0.003$ and $\alpha_{\ssh}=20.6\pm0.2$ 
(see details in Supplementary Materials). Hence, in the thermodynamic limit, we expect the probability of the two edge states to remain non-vanishing, with $P_{\ssh}>0$.

We then fix $a=-1$ and the system size $L$, and analyze how $P$ changes with the bond dissipation strength 
$\Gamma_0$ and the inter-cell hopping amplitude $w$, as shown in Fig.~\ref{fig1}(c,d). We observe that $P$ increases as 
$\Gamma_0$ decreases, eventually reaching a small approximate plateau near the limit of $\Gamma_0\to0^+$ [Fig.~\ref{fig1}(c)]. We note that in our numerical calculations, $\Gamma_0$ does not reach $0$ in Fig.~\ref{fig1}(c).
The variation of $P$ with $w$ is more dramatic. Just past the topological transition point, 
$P$ takes a relatively large value, and as $w$ increases, $P$ decreases sharply, and afterward, as 
$w$ continues to increase, $P$ changes only slightly [Fig.~\ref{fig1}(d)]. Therefore, for a finite-size system, a smaller 
$\Gamma_0$ and a value of $w$ closer to $v+0^{+}$ are more favorable for the preparation of topological edge states.

\begin{figure}[ht!]
	\includegraphics[width=0.5\textwidth]{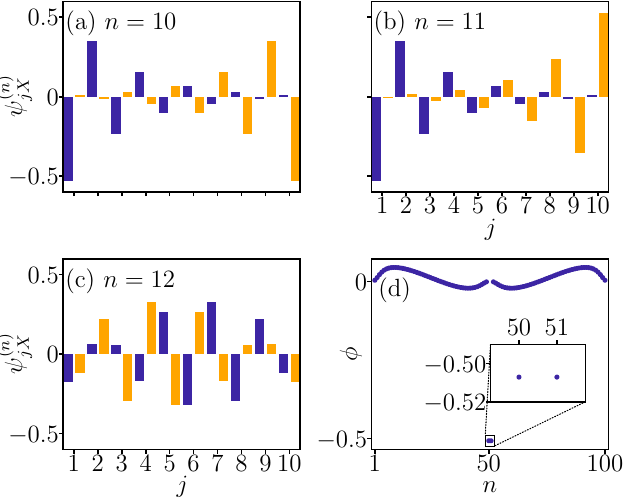}
	\caption{The distributions of (a, b) two edge states and (c) a bulk eigenstate for the SSH model with 
		$N=10$ unit cells. The distribution on the A (B) sublattice is marked in blue (orange), and the horizontal axis $j$ represents the unit cell index.
		(d) The effective relative phase $\phi$ for all eigenstates of $H_\ssh$
		with $L=100$ and $w=1.5v$.
	}
	\label{fig2}
\end{figure}

To understand why bond dissipation can drive the particles in the SSH chain to topological edge states, we first examine the distribution of edge states and bulk states, as shown in Fig.~\ref{fig2}(a-c). One can see that the distribution of the edge states has opposite phases between the same species of sublattice sites in nearest-neighbor cells [Fig.~\ref{fig2}(a,b)]. Such a pattern is absent in the bulk state of $H_{\ssh}$ [Fig.~\ref{fig2}(c)]. Bond dissipation Eq. \eqref{eq_ssh_diss_samesub} with 
$a=-1$ is designed to change the phase distribution of the same species of 
sublattice sites in nearest neighboring cells from the in-phase to an out-of-phase, thus it can prompt the system to favor the topological edge states. To further gain a quantitative understanding, we introduce a quantity $\phi$ that contains both the phase information of the eigenstate distribution of $H_\ssh$ and the distribution of the dissipation operator.
Since $H_\ssh$ is real and symmetric, the matrix elements of its eigenstates are also real. For the $n$-th eigenstate
$\ket{\psi^{(n)}}$, the sign of $\psi_{jX}^{(n)}\psi_{(j+1)X}^{(n)}$ reflects whether the wave function at points $(j,X)$ and $(j+1,X)$ is in-phase or out-of-phase. Here, $\psi_{jX}^{(n)}$ represents the projection of the state $\ket{\psi^{(n)}}$
 onto the sublattice $X$ (where $X=A, B$) of the 
$j$-th unit cell, i.e.,  $\psi_{jX}^{(n)}=\braket{j,X|\psi^{(n)}}$.
Furthermore, the final form of the steady state depends not only on $\psi_{jX}^{(n)}\psi_{(j+1)X}^{(n)}$ but also on the corresponding dissipation strength between $(j,X)$ and $(j+1,X)$. For example, if the dissipation strength between $(1,A)$ and $(2,A)$ 
tends to infinity, while the dissipation strength elsewhere is very small, then the final steady state will clearly depend only on the phase information between the state at $(1,A)$ and $(2,A)$. Based on the above discussions, we introduce 
$\phi$ as
\begin{equation}
	\phi
	%=\sum_{X=A,B}\sum_{m}\phi_{m(m+\ell)}^{(X)}
	=\sum_{X=A,B}\sum_{j}\psi_{jX}^{(n)}\psi_{(j+1)X}^{(n)}\sqrt{\Gamma_{jX}/\Gamma_0}
	\label{eq_relPhase_ssh}
\end{equation}
to quantify the effective relative phase of $\ket{\psi^{(n)}}$ between sites on the same sublattice of neighboring unit cells. Here, the appearance of $\Gamma_0$ in \eqref{eq_relPhase_ssh} serves to make $\phi$
dimensionless. Fig.~\ref{fig2}(d) shows that $\phi$ is negative and has the largest absolute value in the two edge states of 
$H_\ssh$. This is why bond dissipation Eq. \eqref{eq_ssh_diss_samesub} with $a=-1$ can drive the SSH chain to the edge states. \\

\noindent\textbf{Kitaev chain}\\
 Majorana fermions, proposed as a key component of topological quantum computers, have attracted enormous interest over the past two decades (see Refs.~\cite{MZMREF1,MZMREF2,MZMREF3,MZMREF4,MZMREF5,MZMREF6} for reviews). The pioneering paper by Kitaev suggests that Majorana fermions can manifest as MZMs in a spinless p-wave superconductor~\cite{Kitaev}. Our second example is the Kitaev chain, described by 
\begin{equation}
	H_\kc=\sum_{j}(-tc_{j+1}^\dagger{}c_j
	+\Delta{}c_{j+1}c_j+h.c.)-\mu\sum_jc_j^\dagger{}c_j,
\end{equation}
where $t$, $\Delta$ and $\mu$ represent the hopping amplitude, superconducting
pairing gap, and chemical potential, respectively.
When $\mu<2t$, the lowest-lying eigenstates of $H_\kc$ are topological 
and degenerate in the thermodynamic 
limit, with one state in the even-particle subspace and the other in the odd-particle subspace. This degeneracy arises because $H_\kc$ commutes with the parity operator $\Pi=(-1)^{\sum_jc_j^\dagger{}c_j}$, which conserves fermion parity.  

\begin{figure}[ht!]
	\centering
	\includegraphics[width=0.45\textwidth]{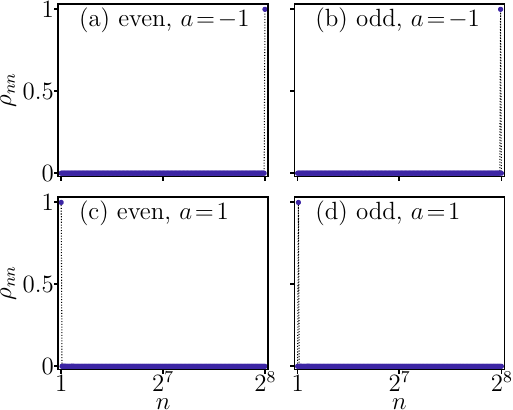}
	\caption{Steady states of Kitaev chain under bond dissipation. 
	The horizontal axis represents the index of the eigenstate $\ket{\Psi^{(n)}}$ in ascending order of the corresponding eigenenergy. $\rho_{nn} = \braket{\Psi^{(n)}|\rho_s|\Psi^{(n)}}$ is the probability of obtaining the $n$th eigenstate.} 
	\label{fig3}
\end{figure}

\begin{figure}[ht!]
	\centering
	\includegraphics[width=0.5\textwidth]{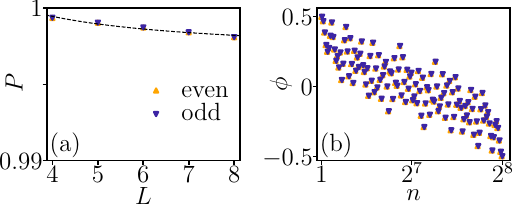}
	\caption{(a) The probability $P$ of the ground states in the steady state for different system sizes, with the black dashed line representing a linear fit of $P$ as a function of $1/L$. (b) The relative phase $\phi$ of each eigenstate of $H_\kc$. $P$ and $\phi$ are represented by blue inverted triangles in the odd-particle subspace and orange triangles in the even-particle subspace. Here we fix $a=1$, $L=8$, $\mu=0.1t$ and $\Delta=t$.
	} 
	\label{fig4}
\end{figure}

We consider the bond dissipation only at the ends of the Kitaev chain, where $\Gamma_j=\Gamma_0$ for $j=1, L-1$, and $\Gamma_j=0$ for all other sites $j\neq1, L-1$ in Eq. \eqref{eq_bond_diss_jump}. Since the bond dissipation operator does not alter the particle number, the Lindbladian $\mathcal{L}$ can be diagonalized separately in the even and odd subspaces, leading to two types of steady states corresponding to even and odd parity.
In the numerical simulation, we set $L=8$, $\mu=0.1t$ and $\Delta=t$ for the Hamiltonian $H_\kc$, and 
$\Gamma_0=10^{-4}t$ and $\ell=1$ for the bond dissipation.
Our results show that,  for both even and odd particle subspaces, bond dissipation with 
$a=-1$ drives the system to its highest excited states, as shown in Fig.~\ref{fig3}(a,b). In contrast, when 
$a=1$, bond dissipation drives the system to the ground state, as shown in Fig. 3(c,d). The steady-state density matrix 
occupies the ground state with a probability of up to $P=0.998$ in both even and odd subspaces for 
$a=1$. We fix $a=1$ and fit the ground state occupation probability $P$ as a function of system size $L$, as shown in Fig.~\ref{fig4}(a). 
$P$ can be linearly fitted with respect to $1/L$ as $P=P_\kc+\alpha_\kc/L$, with numerical calculations yielding
$P_\kc=0.9970\pm0.0002$ and $\alpha_\kc=0.010\pm0.001$. In the thermodynamic limit $L\to\infty$,
$P\to{}P_\kc$ which retains a very high value. Therefore, by utilizing bond dissipation, it is possible to assist in preparing the system to the ground state, thereby facilitating the creation of MZMs.

To understand such an efficient preparation effect, we examine 
the eigenstates of $H_\kc$. The $n$th eigenstate $\ket{\Psi^{(n)}}$ can be expressed as 
\begin{equation}
	\ket{\Psi^{(n)}} 
	= (\Psi_{0}^{(n)}
	+\sum_{\beta=1}^L\sum_{j_\beta>\ldots>j_1}
	\Psi^{(n)}_{j_\beta\ldots{}j_1}c_{j_\beta}^\dagger\ldots{}c_{j_1}^\dagger)
	\ket{\varnothing},
\end{equation}
where $\Psi_{0}^{(n)}$ is the projection of the wave function onto the vacuum state, $\beta$ represents the total number of particles, and $j_1\ldots j_{\beta}$ denote the lattice positions occupied by the $\beta$ particles. Since bond dissipation only acts on the two ends of the chain, according to Eq. \eqref{eq_relPhase_ssh}, we only need to consider the relative phases of the wavefunction between the first and second lattice sites, as well as between the $(L-1)$-th and $L$-th lattice sites. Since these two relative phases are equal, we only need to consider the relative phase of the wavefunction between the first two lattice sites and multiply it by $2$, namely (see Methods): 
\begin{equation}
	\phi=2\left(\Psi_1^{(n)}\Psi_2^{(n)}+\sum_{\beta=2}^{L-1}\sum_{j_\beta>\ldots>j_2>2}\Psi_{j_\beta\ldots j_2{}1}^{(n)}\Psi_{j_\beta\ldots j_2{}2}^{(n)}\right).
	\label{eq_relphase_kc}
\end{equation}
In the first term, $\Psi_1^{(n)}$ and $\Psi_2^{(n)}$ represent the projections of the wavefunction $\Psi^{(n)}$ onto the first and second lattice sites, respectively, when the system contains only one particle. In the second term, $\Psi_{j_\beta\ldots j_2{}1}^{(n)}$ and $\Psi_{j_\beta\ldots j_2{}2}^{(n)}$ represent the same occupation on lattice sites greater than or equal to $3$, but the former corresponds to the particle occupying the first lattice site (with $j_\beta>\ldots>j_2>2$) and the second site unoccupied, while the latter corresponds to the particle occupying the second lattice site with the first unoccupied (see Methods). As shown in Fig.~\ref{fig4}(b), the ground state has the most positive relative phase (in-phase), while the highest excited state has the most negative relative phase (out-of-phase). In the Supplementary Material, we also calculated the darkness describing the dissipation effect on each eigenstate and obtained results completely analogous to those in Fig.~\ref{fig4}(b). Therefore, bond dissipation with $a=1$ and 
$a=-1$ drives the system towards the ground state and the highest excited state, respectively, as shown in Fig.~\ref{fig3}.\\

\noindent{\large{\textbf{Discussion}}}

\noindent We have demonstrated that bond dissipation in topological systems can lead to a significant proportion of boundary states in the steady state. After the system reaches steady state, removing the dissipation allows us to easily obtain the time evolution of the density matrix: $\rho(t)=e^{iHt}\rho e^{-iHt}=\sum_{mn}e^{iE_mt}|\psi_m\rangle\langle \psi_m|\rho|\psi_n\rangle\langle \psi_n|e^{-iE_nt}=\sum_{mn}e^{i(E_m-E_n)t}\rho_{mn}|\psi_m\rangle\langle \psi_n|$, where $\rho_{mn}$ represents the density matrix in the eigenbasis of the Hamiltonian. It is easy to see that for the diagonal elements with $m=n$, $\rho_{mm}(t)=\rho_{mm}$, meaning the diagonal elements do not change with time. Therefore, dissipation can be used to help prepare topological boundary states, especially in the case of the Kitaev chain, where dissipation drives the system to the ground state, which is beneficial for the preparation of MZMs. We explain this using the phase characteristics of the wave function distribution on neighboring lattice sites. When disorder is introduced, it localizes the bulk without affecting the bond dissipation-assisted preparation of topological boundary states, which we discuss in the Supplementary Material using the SSH model as an example. Our work also raises some interesting questions worth exploring, such as: what would be the effects of bond dissipation on topological systems if extended to two-dimensional or three-dimensional systems, or even higher-order topological insulators?\\

\noindent{\large{\textbf{Methods}}}\\

\textbf{Calculate the steady state}\\
To numerically calculate the steady state, we need a vector representation
of the density matrix $\rho$ and a matrix representation of the Liouvillian superoperator $\mathcal{L}$. 
Under the computational basis $\{\ket{i}\}$, 
$\rho=\sum_{i,j}\rho_{ij}\ket{i}\bra{j}$, and this density matrix can be mapped to a vector
\begin{equation}
	\ket{\rho} = \sum_{i,j}\rho_{ij}\ket{i}\otimes\ket{j}.
	\label{eq_rho_vectorized}
\end{equation}
%Here we use $\kket{\bullet}$ to signify that it is not a conventional wave function.
And consequently, we can express the Liouvillian superoperator $\mathcal{L}$ as a matrix
\begin{eqnarray}
	\mathcal{L} 
	&=&\sum_j\left[{D}_j\otimes{D}_j^*
	-\frac{1}{2}\left({D}_j^\dagger{D}_j\otimes\mathbbm{1}
	-\mathbbm{1}\otimes{D}_j^\mathrm{T}{D}_j^*\right)\right]
	\nonumber\\
	&&-i\left({H}\otimes\mathbbm{1}-\mathbbm{1}\otimes{H}^\mathrm{T}\right).
	\label{eq_Lmatrix}
\end{eqnarray}
We diagonalize the operator  $\mathcal{L}$ and obtain its eigenvalues and corresponding eigenvectors
 $\ket{\rho}$. The eigenvalue equal to $0$ corresponds to the steady state $\ket{\rho_s}$,  from which we can derive the steady state $\rho_s$ by using \eqref{eq_rho_vectorized}. 

Given the dimension of the Hamiltonian's Hilbert space as $D_H$, it follows from Eq. \eqref{eq_Lmatrix} that
$\mathcal{L}$ is a $D_H^2\times{}D_H^2$ matrix. For the SSH model, if the system size is 
$L=100$, meaning $D_H=100$, then $\mathcal{L}$ is a $10^4\times10^4$ matrix. In the case of the Kitaev chain, if the size is $L=8$, and considering that particle number is not conserved, $D_H=2^L=2^8$, so $\mathcal{L}$ is a  $2^{16}\times{}2^{16}$ matrix. However, since the parity $\Pi$ is conserved, we can diagonalize $\mathcal{L}$
separately within the odd-particle and even-particle subspaces. Within each subspace, the matrix representation of
$\mathcal{L}$ is $2^{14}\times{}2^{14}$, as both the odd-particle and even-particle subspaces have dimension
 $D_H=2^7$.\\

\textbf{The relative phase of neighboring lattice sites in a system of many particles}\\

We use Eq. \eqref{eq_relphase_kc} to describe the relative phase of the state $\Psi^{(n)}$ of the Kitaev chain between lattice sites $1$ and $2$. The first term corresponds to the product of the projections of the wavefunction onto the two lattice sites in the single-particle case. Now, let us explain the second term, i.e., $\sum_{\beta=2}^{L-1}\sum_{j_\beta>\ldots>j_2>2}\Psi_{j_\beta\ldots j_2{}1}^{(n)}\Psi_{j_\beta\ldots j_2{}2}^{(n)}$.
If neither the first nor the second lattice site is occupied by a particle, then $D_1 = \sqrt{\Gamma_1}(c_{j+1}^\dagger + a c_j^\dagger)(c_{j+1} - a c_j)$ clearly gives zero when acting on the system. If both the first and second lattice sites are occupied by particles, it can also be shown that the action of $D_1$ results in zero. The proof is as follows:
$(c_2^\dagger+ac_1^\dagger)(c_2-ac_1)c_2^\dagger{}c_1^\dagger\ket{\varnothing}
=(c_2^\dagger+ac_1^\dagger)(c_1^\dagger+ac_2^\dagger)\ket{\varnothing}
=c_2^\dagger{}c_1^\dagger-a^2c_2^\dagger{}c_1^\dagger\ket{\varnothing}
=0$.  Here, we use the condition $a=\pm 1$. This is why the total number of particles, $\beta$, only needs to be calculated up to $L-1$. If calculated up to $L$, it would imply that both the first and second lattice sites are occupied by particles.
Therefore, we only consider the phase difference between the particle occupying the first lattice site and the particle occupying the second lattice site (for occupations on lattice sites greater than or equal to $3$, the occupation is the same). Through a linear transformation, the state in this case can also be written in the basis of the in-phase and out-of-phase states formed by the first two lattice sites.\\

%\noindent {\large{\textbf{Data availability}}}\\
%The data used in this study are available in the GitHub repository \href{https://github.com/G-CX1/STL-Code}{https://github.com/G-CX1/STL-Code}.\\

%\noindent {\large{\textbf{Code availability}}}\\
%The code used in this study is available in the GitHub repository \href{https://github.com/G-CX1/STL-Code}{https://github.com/G-CX1/STL-Code}.\\

\noindent {\large{

\vspace{0.5cm}

\begin{acknowledgments}
\noindent {\large{\textbf{Acknowledgments}}}\\
This work is supported by National Key R\&D Program of China (Grant No. 2022YFA1405800 and Grant No. 2023YFA1406704), the Key-Area Research and Development Program of Guangdong Province (Grant No.2018B030326001), Guangdong Provincial Key Laboratory(Grant No.2019B121203002). Y. P. acknowledges support from the Research Fund of Post-doctor who came to Shenzhen (Grant No. K202402020309001). H. H. is also supported by the National Natural Science Foundation of China (Grant No.
12474496).\\
\end{acknowledgments}

\noindent {\large{\textbf{Author contributions}}}\\
Y.W. conceived the idea, performed the theoretical analysis with contributions from H.H., Y.P.and C.Y. Y.W. and Y.P. wrote the major part of the paper. Y.P. did the calculations and wrote the code for analysing. C.Y., H.H. and Y.W. participated in the discussion of the numerical calculations.\\

\noindent {\large{\textbf{Competing interests}}}\\
The authors declare no competing interests.\\

\clearpage
\setcounter{equation}{0}
\setcounter{figure}{0}
\renewcommand{\thefigure}{S\arabic{figure}}
\renewcommand{\theequation}{S\arabic{equation}}

\setcounter{tocdepth}{5}
\setcounter{secnumdepth}{5}
\renewcommand{\thesection}{S\arabic{section}}
\renewcommand{\thesubsection}{\thesection.\Alph{subsection}}

\onecolumngrid
%\newpage
%\renewcommand{\theequation}{S\arabic{equation}}
%\renewcommand{\thefigure}{S\arabic{figure}}
%\renewcommand{\thetable}{S\arabic{table}}
%\setcounter{equation}{0}
%\setcounter{figure}{0}
%\setcounter{table}{0}
%%%%%%%%%%%%%%%%%%%%%%%%%%%%%%%%%%%%%%%%%%%%%%%%%%%%%%%%%%%%%%%%%%%%

\begin{center}
    {\bf \large Supplementary Material for ``Dissipative-assisted preparation of topological boundary states'' }
\end{center}

\noindent This supplementary material provides details on:\\
\noindent (I) Size-scaling behavior of $P$ for the SSH model;\\
\noindent (II) Other types of dissipation strength configurations;\\
\noindent (III) SSH model with random intra-cell hopping;\\
\noindent (IV) Topological states of the Kitaev chain;\\
\noindent (V) Darkness of eigenstates of the Kitaev chain.\\

\section*{(I) Size-scaling behavior of $P$ for the SSH model}
We fit the occupation probability $P$ of edge states to
the system size $L$ for the steady states of the SSH under bond dissipation with $a=-1$. We use polynomial fitting of $P$ as a function of the inverse system size $1/L$. 
The explicit fitting formula is 
\begin{equation}
	P=P_\ssh+\alpha_\ssh/L.
	\label{eq_ssh_P_fit}
\end{equation}
It can be seen from Fig.~\ref{figs1} that this fit is almost perfect for the numerical data obtained. 
The fitting coefficients are $P_\ssh=0.131\pm0.003$ and $\alpha_\ssh=20.6\pm0.2$.
\begin{figure}[ht!]
	\centering
	\includegraphics[width=0.309\textwidth]{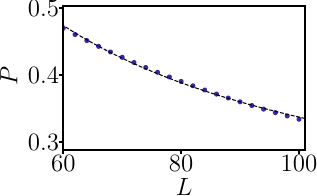} 
	\caption{Size-scaling of the occupation probability $P$ of edge states in the SSH model. 
		The blue dots represent $P$ calculated numerically, and the black dashed line is the fitting curve 
		\eqref{eq_ssh_P_fit}. Here $a=-1$, $w=1.5v$ and $\Gamma_0=0.06v$.} 
	\label{figs1} 
\end{figure}

\section*{(II) Other types of dissipation strength configurations}
In the main text, we discussed that the strength of bond dissipation applied to the SSH model decays exponentially from the edges to the bulk. Here, we discuss other types of dissipation strength configurations. The most trivial one is a homogeneous configuration, where $\Gamma_{jX}$ in Eq. \eqref{eq_ssh_diss_samesub} is the same everywhere, i.e., $\Gamma_{jX} = \Gamma_0$. As shown in Fig.~\ref{figs2}(a), such a bond dissipation with $a = -1$ and $\ell = 2$ can only drive the system towards two bulk states near the gap.

Another type of configuration we consider is given by 
\begin{equation}
	\Gamma_{jA} = \Gamma_0/j^z,
	\quad
	\Gamma_{jB} = \Gamma_0/(N-j)^z,
	\label{eq_bond_diss_strength_invp}
\end{equation}
Where $z>0$. The bond dissipation strength $\Gamma_{jX}$ is 
concentrated at the two ends of the SSH chain, with the degree of concentration
determined by the value of $z$. For $a=-1$
and $\ell=2$, the SSH chain is driven to: (i) two
bulk states when $z=1$ (Fig~\ref{figs2}(b)); (ii) the 
two edge states, plus two
bulk states nearest to them, within the ordered energy levels of 
$H_\ssh$ when $z=1.5$ (Fig~\ref{figs2}(c)); and (iii) 
the two edge states when $z=2$
(Fig~\ref{figs2}(d)). Therefore, bond dissipation with
configuration~\eqref{eq_bond_diss_strength_invp} can drive
the SSH chain to edge states when $z\ge2$.
\begin{figure}[ht!] 
	\centering
	\includegraphics[width=0.5\textwidth]{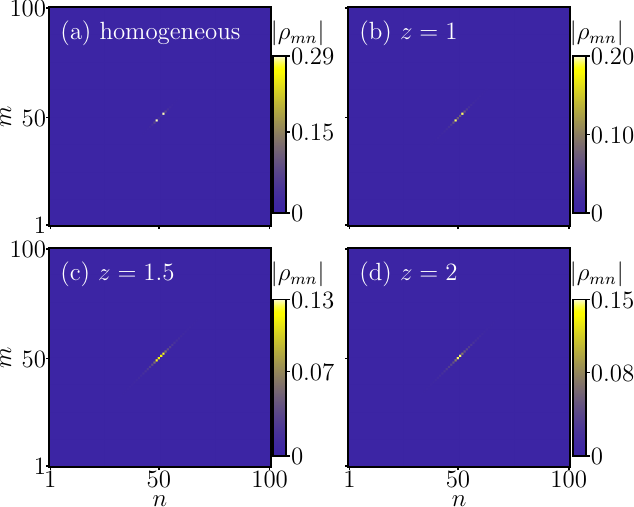} 
	\caption{Steady-state density matrix $\rho_{mn}$ for different dissipation strength 
		configurations. 
		(a) $\Gamma_j$ is homogeneous. The dissipation configuration of type \eqref{eq_bond_diss_strength_invp} 
		with (b) $z=1$, (c) $z=1.5$ and (d) $z=2$. Here $L=100$, $w=1.5v$, $\Gamma_0=0.06v$, $a=-1$ and $\ell=2$.} 
	\label{figs2} 
\end{figure}

\section*{(III) SSH model with random intra-cell hopping}

	\begin{figure}[ht!] 
	\centering
	\includegraphics[width=0.45\textwidth]{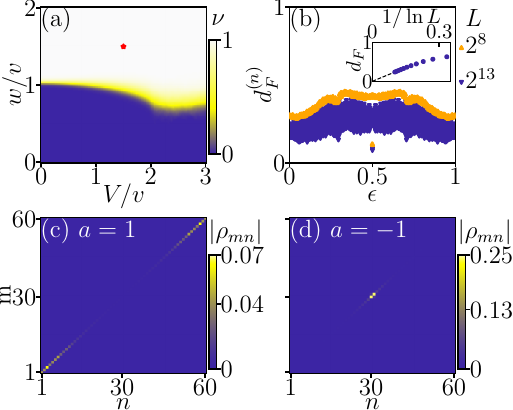} 
	\caption{(a) Phase diagram of SSH chain under random intra-cell hopping.
		(b) Fraction dimension for the eigenstates of $\tH_\ssh$. 
		All fraction dimension $d_F^{(n)}$ are calculated from the $IPR$ averaged over $20480/N$ samples of 
		random intra-cell hopping. Blue inverted triangles represent $d_F^{(n)}$ for a system 
		size of $L=2^{13}$, and orange triangles represent $d_F^{(n)}$ for $L=2^8$.
		The inset panel shows the average $d_F$ (blue dots) of $d_F^{(n)}$ over all eigenstates of 
		$\tH_\ssh$, where the dashed black line represents the linear fit of $d_F$ in 
		the panel as a function of $\ln{}L$.
		Steady states of the SSH model of size $L=60$ with random intra-cell hopping under bond dissipation 
		with (c) $a=1$ and (d) $a=-1$. $w=1.5v$ and $V=1.5v$ are the same for (b), (c) and (d), 
		which is marked as the red star in (a).}
	\label{figs3}
\end{figure}
In this section, we introduce random intra-cell hopping into the $H_\ssh$ model, i.e., 
\begin{equation}
	\tilde{H}_\ssh=H_\ssh + \sum_{j}(V_jc_{jA}^\dagger{}c_{jB}+h.c.),
\end{equation}
where  $V_j$ is a random number uniformly distributed in the range $[-V/2, V/2]$, with $V$ representing the strength of the random intra-cell hopping. The fixed component $v$ of the intra-cell hopping can still be used as the energy unit.
Even with random intra-cell hopping, chiral symmetry is preserved. We can calculate the real-space winding number 
\begin{equation}
	\nu=\frac{1}{N'}\Tr'(S\tH_F[\tH_F,Q]),
\end{equation}
where $N'$ is the number of unit cells chosen at the center of the chain for the trace calculation, 
denoted as $\Tr'$, to avoid boundary effects. In our calculation, we set $N'=N/2$.
$S=\sum_{j}{c}_{jA}^\dagger{}c_{jA}-{c}_{jB}^\dagger{}c_{jB}$ is the chiral operator, 
$\tH_F=(\sum_{m{\le}N}-\sum_{m>N})\ket{\tpsi^{(m)}}\bra{\tpsi^{(m)}}$ is the flat-band version of 
$\tH_\ssh$, and $Q=\sum_{j}j({c}_{jA}^\dagger{}c_{jA}-{c}_{jB}^\dagger{}c_{jB})$ is the coordinate operator.
Here $\ket{\tpsi^{(n)}}$ are the eigenvectors of $\tH_\ssh$ arranged
in ascending order of the corresponding eigenenergies $\tE^{(n)}$.
$\nu=1$ ($\nu=0$) indicates the topological (trivial) phase,  which determines the phase diagram of 
$\tH_\ssh$, as shown in Fig.~\ref{figs3}(a). 	
			
Next, we fix  $w=1.5v$ and $V=1.5v$ (marked as a red star shape in 
Fig.~\ref{figs3}(a)). The system is topological, and there are two edge states
$\ket{\tpsi^{(N)}}$ and $\ket{\tpsi^{(N+1)}}$. Due to the presence of random hopping, the bulk eigenstates become localized. To verify this, we can calculate the fractional dimension $d^{(n)}_F$ of the (normalized) eigenstates $\ket{\tilde{\psi}^{(n)}}$.
\begin{equation}
	d_F^{(n)}=-\ln\braket{IPR^{(n)}}/\ln{}L 
	\quad\textrm{and}\quad
	IPR^{(n)}=\sum_{j=1}^L|\tilde{\psi}_j^{(n)}|^4.
\end{equation}
Here, $IPR$ denotes the inverse participation ratio, and $\braket{IPR}$ represents the average over different random intra-cell hopping samples. As the system size increases, the fractal dimension $d_F$
tends to $1$ ($0$) for the extended (localized) states.
To analyze the behavior of $d_F^{(n)}$ with respect to 
the system size $L$ for the eigenstate $\ket{\tpsi^{(n)}}$,
we assign  an index $\epsilon=(n-1)/(L-1)$ to $\ket{\tpsi^{(n)}}$,  where $\epsilon=0$ corresponds to the ground state and  $\epsilon=1$ corresponds to the highest excited state of $\tH_\ssh$.
As shown in Fig.~\ref{figs3}(b),  $d_F^{(n)}$ decreases
with increasing system size $L$ for every eigenstate $\ket{\tpsi^{(n)}}$. 
We can further extrapolate the finite-size behavior of the average fraction dimension 
$d_F=\sum_nd_F^{(n)}/L$ for all eigenstates of $\tH_\ssh$, 
inferring that $d_F$ approaches $0$ in the thermodynamic limit $L\to\infty$ 
(see the smaller inset panel in Fig.~\ref{figs3}(b)).
Thus, we conclude that all eigenstates of $\tH_\ssh$ are localized. We then introduce bond dissipation \eqref{eq_ssh_diss_samesub} with 
$\Gamma_0=0.06$. This can drive the system to the bulk states at both the bottom and top of the energy spectrum when
 $a=1$ (see Fig.~\ref{figs3}(c)).
On the other hand, bond dissipation with $a=-1$ can drive the system to the two edge states, 
as shown in Fig.~\ref{figs3}(d).
This behavior is similar to that of the clean SSH chain.
We can further engineer particle loss channels on both ends of the chain.
By turning on both particle loss channels and  bond dissipation, we can drain all bulk electrons from the 
system.

\section*{(IV) Topological states of the Kitaev chain}
\begin{figure}[ht!] 
	\centering 
	\includegraphics[width=0.5\textwidth]{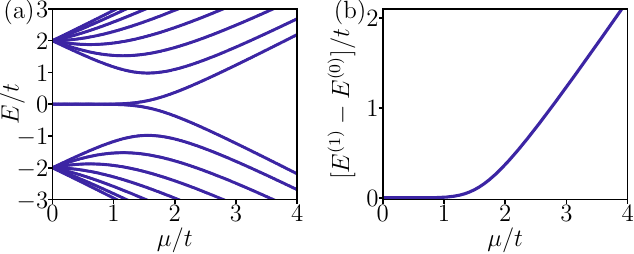} 
	\caption{(a) Energy levels of BdG Hamiltonian and (b) energy difference between the ground state $E^{(0)}$ and the first excited state $E^{(1)}$.    
		Here $L=8$ and $\Delta=t$.}
	\label{figs4}
\end{figure}
The study of the Kitaev chain is typically performed in the Bogoliubov-de-Gennes (BdG) formalism, i.e.,
${H}=\frac{1}{2}\pmb{c}^\dagger\pmb{H}_\BdG\pmb{c}$,
where $\pmb{H}_\BdG$ is called the BdG Hamiltonian which is a 
$2L\times{}2L$ matrix, and $\pmb{c}$ is a $2L\times1$ column vector of annihilation and creation 
operators, $\pmb{c}=({c}_1,\ldots,{c}_L,{c}_1^\dagger,\ldots,{c}_L^\dagger)^{T}$, with  
$\pmb{c}^\dagger$ being the complex conjugate of $\pmb{c}$.
By diagonalizing the BdG Hamiltonian, we obtain a set of Fermionic quasiparticle modes, whose annihilation operators are $f_k$, such that
$H_\kc=\sum_kE_k(f_k^\dagger{}f_k-1/2)$,
where the commutation relations of the fermion modes are satisfied
$\{f_k,f_{k'}\}=0$ {and} $\{f_k,f_{k'}^\dagger\}=\delta_{kk'}$.
$E_k$s are eigenenergies of the $\pmb{H}_\BdG$ in ascending order and they are symmetric with respect 
to $E=0$. When $\mu/t<2$, there are two states in the gap of $\pmb{H}_\BdG$, which are the so-called 
Majorana zero modes (see in Fig.~\ref{figs4}(a)).
The ground state $\ket{\Psi^{(0)}}$ of the Kitaev chain is the vacuum state of the quasiparticle modes, i.e., 
$f_k\ket{\Psi^{(0)}}=0$ for all $k$ and the corresponding energy is $E^{(0)}=-\sum_kE_k/2$. In the topological 
regime $\mu/t<2$, the first excited state 
$\ket{\Psi^{(1)}}$ with eigenvalue $E^{(1)}$ contains one Majorana zero mode. In thermodynamic limit, $\ket{\Psi^{(1)}}$ and $\ket{\Psi^{(0)}}$
are degenerate, $E^{(0)}=E^{(1)}$, as shown in Fig.~\ref{figs4}(b).  According to our calculations in the main text, by diagonalizing the particle-number non-conserving Kitaev chain with a Hilbert space dimension of 
$2^{L}$, the obtained $E^{(0)}$ and $E^{(1)}$ are exactly the same as the results from the BdG formalism. Therefore, our discussion in the main text suggests that bond dissipation can assist in preparing the ground state of the Kitaev chain, which in turn helps in the preparation of Majorana zero modes.

\section*{(V) Darkness of eigenstates of the Kitaev chain}

For an eigenstate $\ket{\Psi^{(n)}}$,  if a set of the dissipative operator 
$D_j$ acts on it and gives zero, i.e., 
$\forall j$ $D_j\ket{\Psi}=0$, then $\ket{\Psi^{(n)}}$
 is a dark state. If there exists a dark state $\ket{\Psi^{(n)}}$, the steady state can be a pure state, $\rho_s=\ket{\Psi^{(n)}}\bra{\Psi^{(n)}}$. For the bond dissipation applied at both boundaries of the Kitaev chain, as mentioned in our main text, we can define the darkness of $\ket{\Psi^{(n)}}$ as 
\begin{equation}
	\mathcal{D}=\left\|D_1\ket{\Psi^{(n)}}\right\|^2+\left\|D_{L-1}\ket{\Psi^{(n)}}\right\|^2
\end{equation}
to quantify how closely $\ket{\Psi^{(n)}}$ resembles a dark state.
In the case where there is no out-of-phase mode at the boundaries of the wavefunction distribution 
 $\ket{\Psi^{(n)}}$, the operators $D_1=(c^{\dagger}_2+ac^{\dagger}_1)(c_2-ac_1)$ with $a=1$ and $D_{L-1}=(c^{\dagger}_L+ac^{\dagger}_{L-1})(c_L-ac_{L-1})$ with $a=1$ act on it and give zero, i.e.,  
$D_1\ket{\Psi^{(n)}}=0$ and $D_{L-1}\ket{\Psi^{(n)}}=0$.  Hence, an eigenstate $\ket{\Psi^{(n)}}$ without out-of-phase modes at both ends of the chain is a dark state under the bond dissipation we discussed.  As shown in Fig.~\ref{figs5}(a), the bond dissipation operators $D_1$ and $D_{L-1}$ with $a=1$ act on the ground state, approximately giving zero. This implies that the ground state is approximately a dark state and also approximately a steady state. When 
$a=-1$, the highest excited state is approximately its steady state (Fig.~\ref{figs5}(b)), which is consistent with the results in the main text. By analogy with Fig.~\ref{figs5} and Fig. 4(b) in the main text, it can be seen that 
 although $\mathcal{D}$ is a very different quantity from the relative phase $\phi$ defined in the main text, they exhibit almost the same shape. Therefore, both $\mathcal{D}$ and $\phi$ can be used to explain why the bond dissipation applied at the boundaries with $a=1$ and $a=-1$ drive the system towards the ground state and the highest excited state, respectively.
\begin{figure}[bt!] 
	\centering 
	\includegraphics[width=0.5\textwidth]{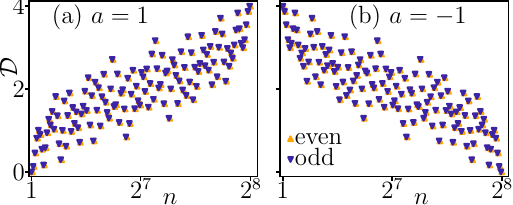} 
	\caption{Darkness of the eigenstate of $H_\kc$ for bond dissipation with phase (a) $a=1$ and 
		(b) $a=-1$. The orange triangle represents the darkness of the eigenstates from the even-particle subspace, and the blue inverted triangle represents the darkness of the eigenstates from the odd-particle subspace.}
	\label{figs5}
\end{figure}

\end{document}